\documentclass[aps,prl,preprint,showpacs,showkeys,groupedaddress]{revtex4}
\newcommand{\W}{14cm}
\usepackage{graphicx}
\usepackage{amsmath}
\begin{document}
\title{Adsorption phenomena in the transport of a colloidal particle through a nanochannel
containing a partially wetting fluid}
\author{German Drazer}
\email{drazer@mailaps.org}
\affiliation{Benjamin Levich Institute and Department of Physics,
City College of the City University of New York, New York, NY 10031}
\author{Boris Khusid}
\email{boris.khusid@njit.edu}
\affiliation{Department of Mechanical Engineering, New Jersey Institute of Technology,
University Heights, Newark, NJJ 07102}
\author{Joel Koplik}
\email{koplik@sci.ccny.cuny.edu}
\affiliation{Benjamin Levich Institute and Department of Physics,
City College of the City University of New York, New York, NY 10031}
\author{Andreas Acrivos}
\email{acrivos@sci.ccny.cuny.edu}
\affiliation{Benjamin Levich Institute and Department of Physics,
City College of the City University of New York, New York, NY 10031}
\date{\today}
\begin{abstract}
Using molecular dynamics simulations, we study the motion of a closely fitting nanometer-size solid 
sphere in a fluid-filled cylindrical nanochannel at low Reynolds numbers and for a wide range of 
fluid-solid interactions corresponding to different wetting situations.
For fluids that are not completely wetting we observe an interesting and novel adsorption phenomenon, 
in which the solid sphere, that was initially moving along the center of the tube, meanders across the channel 
and suddenly adsorbes onto the wall. Thereafter, the adsorbed sphere either {\it sticks} to the
wall and remains motionless on average, or separates slightly from the tube wall and 
then moves parallel to the tube axis, while rotating on average. On the other hand, at short times,
i.e. when the solid particle moves with its center close to the middle of the tube, we find surprisingly 
good agreement between our results and the predictions of the continuum approach 
in spite of the large thermal fluctuations present in our simulations.
\end{abstract}
\maketitle

Recent developments of micro- and nano-fluidic devices for fluid transport
have led to revolutionary new capabilities in the transport and
process of suspended small particles \cite{stone01,giordano01}, with 
the rapid advances in those areas highlighting the crucial importance of
understanding the new features that emerge when 
nanometer size particles move through a fluid-filled nanochannel.
But, in contrast to the extensively studied case of the nano-confinement 
of single fluids which illustrates how their varied hydrodynamic behavior differs
drastically from that observed in macro-scale flows, 
only a few experimental or numerical studies have dealt with the
hydrodynamics of nanometer size particles under similar conditions.

\begin{figure}
\includegraphics*[width=\W]{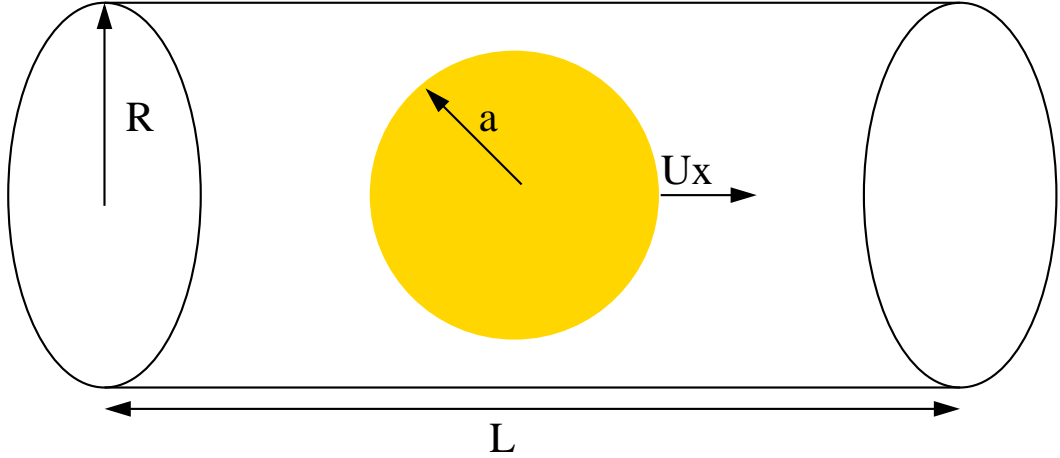}
\caption{\label{system} Schematic view of the colloidal particle of radius $a$
moving through a nanochannel of inner radius $R$.}
\end{figure}

Here, we consider the motion at low Reynolds numbers of a solid sphere in a fluid-filled 
capillary tube when the dimensions of both the particle and the tube approach molecular scales, 
a typical situation in nano-fluidics (see Fig.~\ref{system} for a schematic view of 
the system under consideration). We present the results of molecular dynamics
(MD) simulations, which have already proved very useful in capturing the behavior
of single fluids at fluid-solid interfaces. Motivated by a series of recent 
experimental \cite{cheng02,bonaccurso02,zhu02,tretheway02} 
and numerical \cite{thompson97,sokhan01,cieplak01} studies where the 
wetting properties of the fluid is shown to play a crucial role at
the fluid-solid interface, such as large slip effects for ``nonwetting'' fluids,
we investigate the behavior of the system as the wetting properties 
of the fluid are varied, from perfect to partial wetting.

The hydrodynamic behavior of single fluids in nanochannels and its dependence on the
wetting properties of the liquid, has been reproduced in molecular dynamics (MD)
simulations using simple Lennard-Jones liquids, where the wetting
properties of the fluid were modeled by varying the strength of
the solid-liquid attraction term in the potential \cite{sokhan01,cieplak01}. 
We therefore investigate the motion of a single sphere in a nanochannel
in the simplest possible numerical framework, where all molecular interactions are modeled
by slightly modified Lennard-Jones potentials. 
Our main objective is to investigate to what extent
the molecular character of the system introduces new features not present
when the traditional continuum approach is adopted.

In our MD simulations we consider a tube of inner radius $R$, which is constructed by 
deleting a cylindrical region of atoms from an fcc-lattice. 
The fluid filling the nanochannel has a reduced density 
$\rho \sigma^3=0.8$, where $\sigma$ is the size of the repulsive core, $\sigma \sim O(0.3nm)$, 
and two fluid atoms, separated a distance $r$, interact through the Lennard-Jones potential 
$V_{LJ}(r)=4\epsilon [ (r/\sigma)^{-12}-(r/\sigma)^{-6}]$, truncated 
(and shifted to give a continuous force) at $2.5\sigma$.
The simulations are performed at a temperature $T=1.0 \epsilon/k_B$ (just above the liquid-gas
coexistence value for the bulk Lennard-Jones liquid) and with a characteristic time unit 
$\tau_0 = \sqrt{m\sigma^2/\epsilon}$ (where $m$ is the mass of the fluid atoms), 
$\tau_0$ being $O(1ps)$ for typical liquids. The tube wall has the same density as the fluid, and 
the lattice constant equals $0.85\sigma$, with the wall atoms having mass $100m$ and 
tethered to the fixed lattice sites by a harmonic potential with a large spring constant. 
Following Ref. \cite{koplik89} (see also Ref. \cite{cieplak01} for less dense fluids) 
the wall-fluid interactions are modeled by means of a modified 
Lennard-Jones potential $V_{wf}(r)=4\epsilon [ (r/\sigma)^{-12}- A (r/\sigma)^{-6}]$, 
where the parameter $A$ determines the wetting properties of the fluid-wall system, with the
partially wetting situation corresponding to values of $A$ less than $1$. Specifically,
this molecular model leads to the expression $\cos \theta \approx -1+2A$,  
for the equilibrium contact angle $\theta$ at the fluid-solid interface \cite{barrat99}. Let us mention that, slip
effects at the solid-fluid interface, being negligible at $A=1$, rapidly increase with
decreasing $A$ for this molecular model \cite{koplik89,cieplak01}.
The solid particle is constructed by extracting a sphere of radius $a=5.1\sigma$
from an fcc-lattice having the same lattice constant as the tube wall 
($0.85 \sigma$, $N_A\simeq466$ atoms inside the spherical solid particle and
fixed in their lattice sites). 
The same $V_{wf}$ and $V_{LJ}$ potentials are used for the interactions between 
the atoms of the solid particle and either those of the fluid or those of the tube 
wall respectively. Also, periodic boundary conditions 
are imposed in the direction along the tube axis at $x = L\sim50\sigma$.

\begin{figure}
\includegraphics*[width=\W]{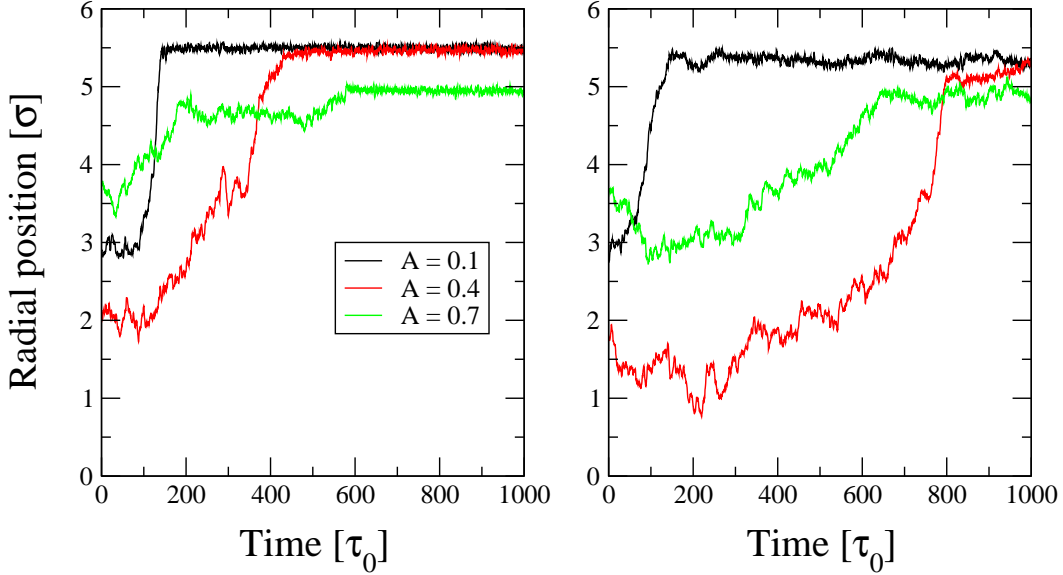}
\caption[radial]{\label{random_motion} Radial position of the solid particle as a function of time for
6 independent MD simulations with $A=0.1$, $0.4$ and $0.7$.
The left graph corresponds to cases in which the particle remains motionless on average (``stick''), 
while on the right graph, for the same values of $A$, we show cases in which the particle is adsorbed 
but moves along the tube axis (``rolling'').}
\end{figure}

We consider first the situation in which a constant force, directed along the axis
of the tube, is applied to the solid particle  
(a force $f=0.1\epsilon/\sigma$ applied to each of its atoms), for a tube radius 
$R=2a$ and various values of the attractive strength $A$, ranging from $0.1$ to $1$. 
In Fig. \ref{random_motion} we present the radial position of the particle as a function of time,
for six different realizations corresponding to $A=0.1$, $0.4$ and $0.7$. In all cases, 
the solid particle not only moves along the tube axis, due to the applied external force,
but it also meanders in the radial direction due to the
thermal fluctuations present in the system. Eventually, 
the particle reaches a critical radial position after which it suddenly 
adsorbes onto the wall and thereafter shows two types of behavior: a {\it stick} situation in 
which the solid particle remains motionless on average, and a {\it rolling} condition in 
which the sphere separates slightly from the tube wall and then moves parallel to the 
tube axis while rotating on average. Moreover, as observed in Fig. \ref{random_motion},
the radial position at which the particle suddenly jumps to the tube wall
increases with $A$. 
It is clear that the potential energy 
reduction due to the adsorption of a solid particle is proportional to $1-A$, 
since the solid particle displaces fluid particles which are less
attracted to both the solid particle and the wall by a factor $A$, and hence the smaller the value of $A$, the
larger the adsorption force. In fact, we find that for $A=1$, where all interactions 
are equal and complete wetting exists, a solid particle initially moving away from the wall does not 
become adsorbed even after very long times when it has meandered all across the tube. On
the other hand, if the solid sphere is initially placed on the tube wall, with
no fluid atoms in the particle-tube gap, the sphere remains adsorbed during the whole duration of the  
simulation $O(1000\tau_0)$. These results indicate that, for $A=1$, a solid
particle moving in the channel is not able to squeeze out all the fluid atoms and become
adsorbed to the tube wall, but, if placed next to the wall, there is no tendency for fluid
atoms to infiltrate between the sphere and the wall, so that the solid 
particle remains adsorbed.

\begin{figure}
\includegraphics*[width=\W]{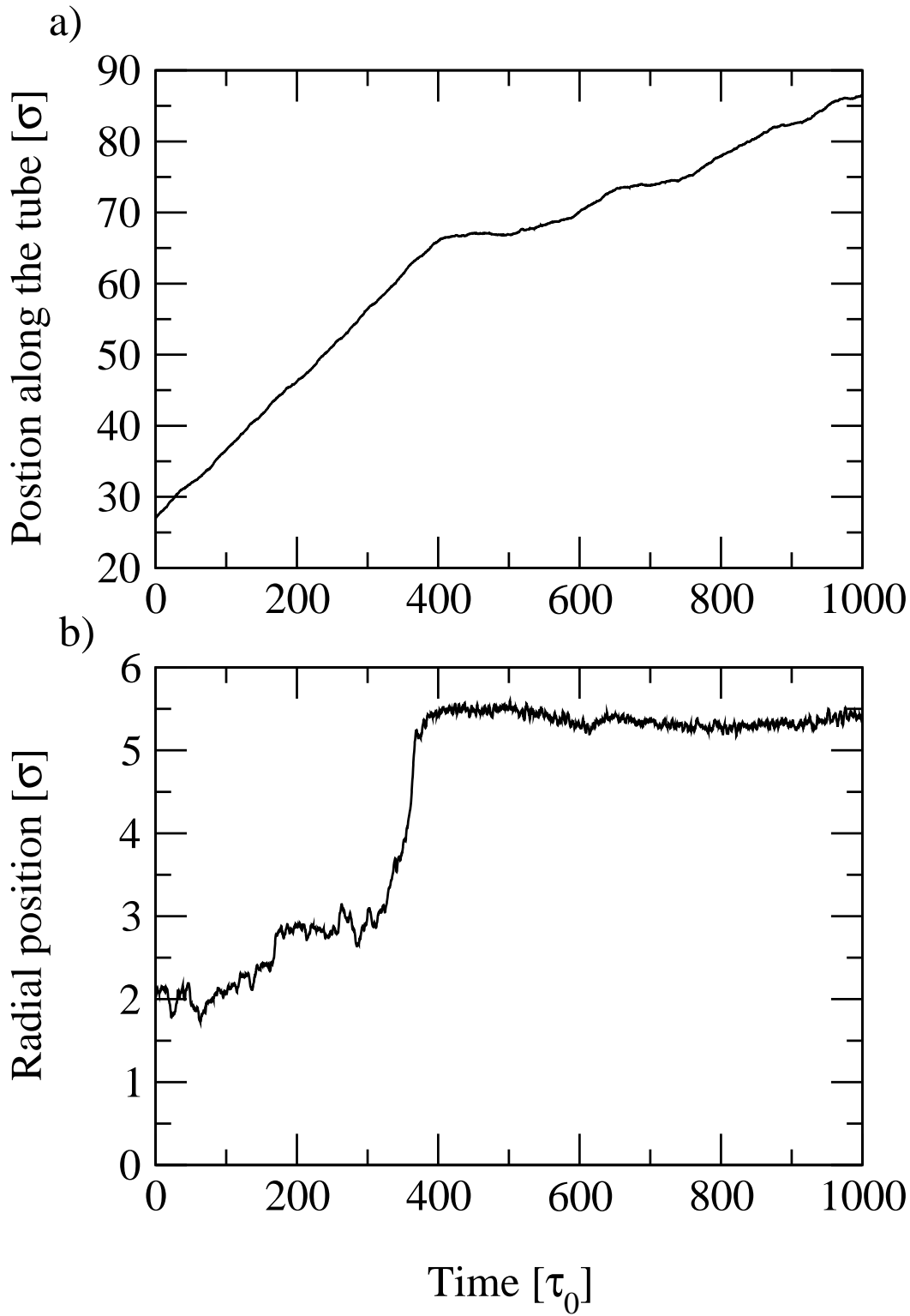}
\caption[motion]{\label{motion} a) Particle position along the tube as a function of time, for
$A=0.1$ and $R=2a$. After the particle is adsorbed ($t\sim400$), alternative ''stick'' 
and ''roll'' situations can be observed. b) Time evolution of the radial position for
the same realization shown in a).} 
\end{figure}

\begin{figure}
\includegraphics*[width=\W]{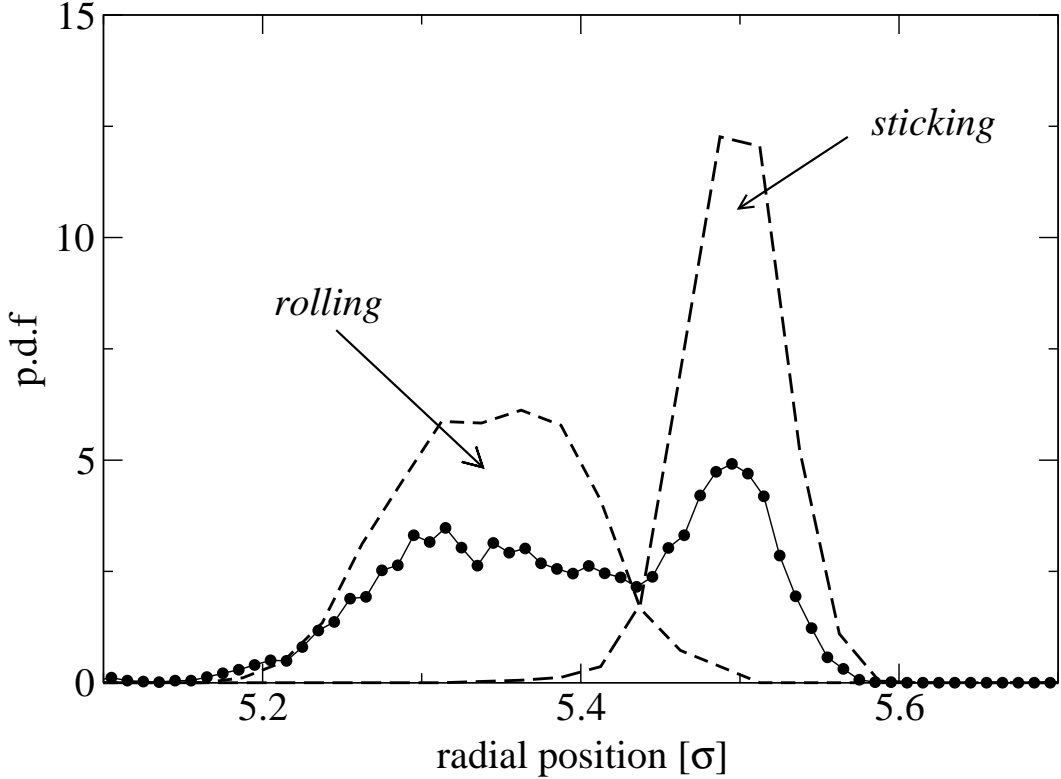}
\caption{\label{pdf} The solid symbols correspond to the probability density function (p.d.f)
for the radial position of the solid particle averaged over 6 different realizations for $A=0.1$. 
Two peaks, corresponding to ``stick'' and ``roll'', can be observed. The dashed lines refer
to the p.d.f's of the two realizations shown in Fig.~\ref{random_motion} for $A=0.1$, 
where the particle is either motionless or rolling along the tube wall.}
\end{figure}

\begin{figure}
\includegraphics*[width=\W]{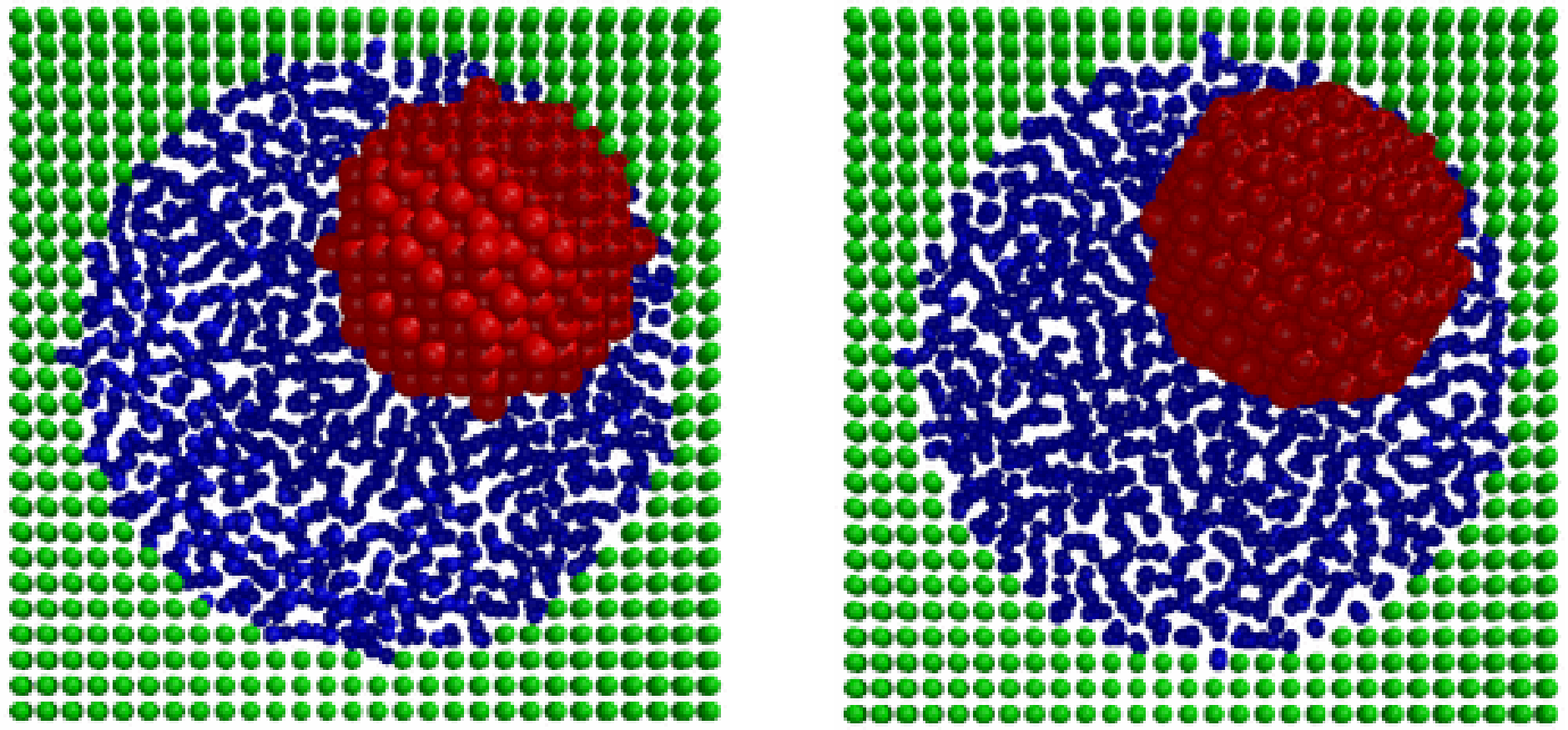}
\caption[snaphots]{\label{snapshot} Snapshots of the solid particle adsorbed to the wall
for the case $A=0.7$ shown in Fig.~\ref{random_motion}. On the left-side snapshot the sphere is
motionless whereas on the right-side the particle is rolling along the tube. 
In both cases, however, the gap between the sphere and the wall is smaller 
than the size of a single fluid atom}
\end{figure}

The two different regimes that are observed after the particle has been adsorbed at the wall 
are also shown in Fig.~\ref{random_motion} where, on the left-side, we present 
cases in which the particle remains motionless on average, but where fluctuations in its radial 
position can be observed. On the other hand,  the right-side 
of Fig.~\ref{random_motion} shows examples
where the particle continues to roll after being adsorbed, and exhibits much larger fluctuations
in its radial position (e.g. compare the two $A=0.1$ cases shown in Fig.~\ref{random_motion}
for times $t > 200 \tau_0$). In some cases, we also observe
an intermittent behavior between the two regimes, as can be seen in Fig. \ref{motion}.
Figure \ref{pdf} gives the probability density function (p.d.f) for
the radial position of the solid particle once it has been adsorbed to the wall, where
the subtle difference between the ``stick'' and ``roll'' regimes is evident.
It is interesting to note that, when the particle is adsorbed, 
particularly for low values of $A$ and in the ``stick'' regime, almost all the fluid atoms
have been squeezed out from the particle-wall gap (see Fig. \ref{snapshot}), a phenomenon which would 
have required an infinite force in the continuum limit \cite{bungay73}. The absence of liquid
atoms in the particle-wall gap results in a pressure imbalance that causes the particle
to remain adsorbed (note that this pressure imbalance plays a particularly important role 
in the case $A=1$, in that, with all interactions being alike, it is the only 
cause of adsorption for a particle placed next to the tube wall). 
This situation strongly resembles the phenomenon of capillary drying on nanometer
length scales observed experimentally \cite{lum99}, in which, when a fluid-filled 
gap between two hydrophobic surfaces falls below a critical value, the liquid is spontaneously
ejected from the gap.

\begin{figure}
\includegraphics*[width=\W]{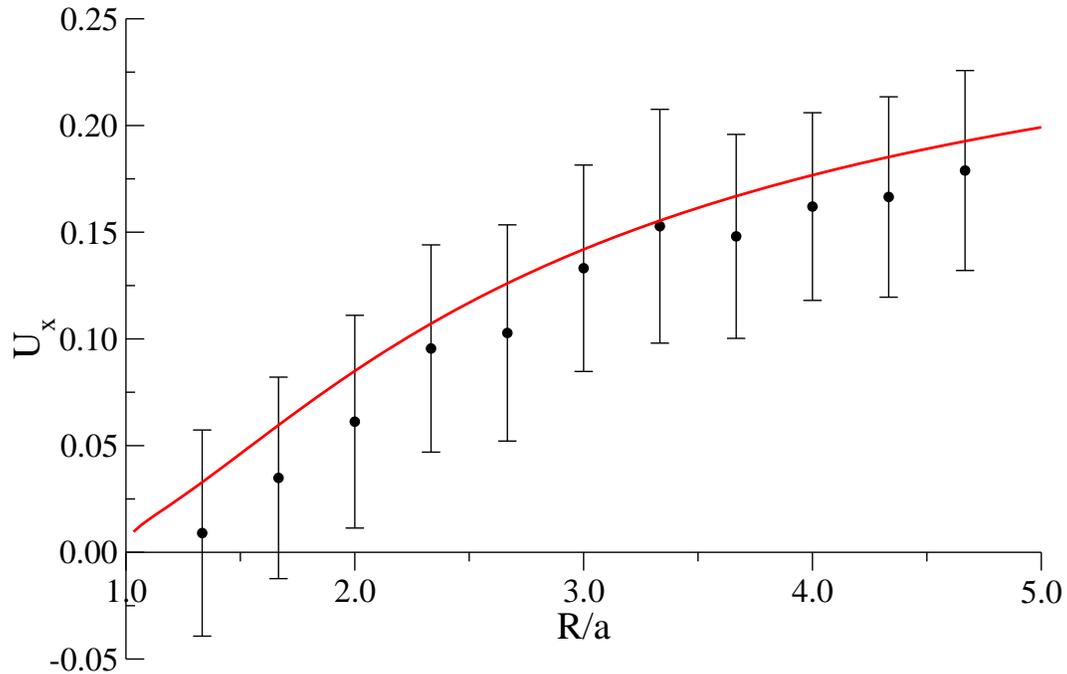}
\caption{\label{bungay} Mean particle velocity along a tube at short times, when a constant force is
acting on the sphere and for $A=1$. Points correspond to MD simulations and the solid 
line corresponds to the continuum results of Bungay and Brenner \cite{bungay73}.
The error bars correspond to temporal fluctuations in the instantaneous velocity
of the solid particle.}
\end{figure}

In order to investigate the effects of varying the ratio of the tube radius to that of the
sphere, $R/a$, we have also performed MD simulations of the same sphere, initially
placed at the center of the tube, moving through the tube
for different values of $R$, ranging from $7\sigma$ to $24\sigma$. First, 
we investigated the case of $A=1$, where slip effects
at the solid-liquid interface have been shown to be negligible even at the molecular level \cite{cieplak01}, 
and which also corresponds to complete wetting.
In Fig. \ref{bungay} we show the results thus obtained for the mean particle 
velocity along the tube axis, $U_x$, as a function of the relative radius, $R/a$, when a constant
force is applied to the sphere ($f=0.1\epsilon/\sigma$, as before). 
The results are for short times, in that the averaging time is too small for
the particle to have meandered away from the center of the tube towards the tube wall.
For purposes of comparison, we also present
the continuum solution for a non-Brownian, perfectly smooth sphere, moving
along the center of the tube in the limit of vanishingly small inertia effects \cite{bungay73}.
It can be seen that, despite the large thermal fluctuations present in the system
and the fact that the Reynolds number in our simulations is not exactly zero ($\text{Re}\lesssim0.2$), 
the continuum calculations are in good agreement with the reduction observed in the mobility of the sphere 
as the tube radius decreases.
Moreover, the fact that the continuum solution slightly overestimates the mean velocity along 
the tube might simply be due to the transverse random motion of the particle in the MD simulations.
We have also investigated the effects of the wetting properties
of the fluid, and find that, although the average sphere velocity increases with decreasing values of $A$,
deviations from the continuum results using the no-slip boundary conditions are small even for attraction 
strengths as small as $A=0.05$ (the deviations in the average particle velocity
from that predicted by continuum hydrodynamics are within 30\%). 
Let us remark that the simulation results are for short times, in that the solid particle remains
close to the center of the tube and therefore adsorption effects are absent even for the smallest
values of $A$.
Similar conclusions concerning the robustness
of continuum calculations for a sphere approaching a plane wall were presented in Ref. \cite{vergeles96}

In summary, an interesting adsorption phenomenon was identified in the transport of
a colloidal particle suspended in a partially wetting fluid though a nanochannel. We also
observed ``stick'' and ``roll'' regimes for the adsorbed particle, in which the ``stick''
case apparently corresponds to a particular orientation of the sphere roughness with respect
to the molecular roughness of the tube wall.
Surprisingly, all fluid atoms are displaced out from the particle-wall
gap in both regimes, something which would have required an infinite force in the
 traditional continuum description.
Finally, in view of what has been presented above, it is clear that understanding this 
wetting-induced adsorption phenomenon is crucial to the successful modeling of the 
transport of colloidal particles in nanofluidic devices.

A.A. and G.D. were partially supported by the Engineering Research Program, Office of Basic Energy Sciences, 
U. S. Department of Energy under Grant DE-FG02-90ER14139;
J.K. was supported by NASA's Office of Biological and Physical Research; 
B.K. was partially supported by a grant from the Defense Advanced Research Project Agency;
G. D. was partially supported by CONICET Argentina.


\end{document}